\documentclass[prb,a4paper,twocolumn,footinbib,showpacs]{revtex4}

\usepackage[latin1]{inputenc}
\usepackage{amssymb}
\usepackage{amsmath}
\usepackage{amsbsy}
\usepackage{bm}
\usepackage{graphicx}

\begin{document}

\title{Nonlinear evolution of the step meandering instability  of a growing crystal surface}
\author{Thomas Frisch}
   \altaffiliation{\'Ecole Générale des Ingénieurs de Marseille,
   Technopôle de Château-Gombert,
   Marseille, France}\email{frisch@irphe.univ-mrs.fr}
   \affiliation{%
   Institut de Recherche sur les Phénomènes Hors \'Equilibre,
   UMR 6594, CNRS, Université de Provence, Marseille, France}

   \author{Alberto Verga}\email{Alberto.Verga@irphe.univ-mrs.fr}
   \affiliation{%
   Institut de Recherche sur les Phénomènes Hors \'Equilibre,
   UMR 6594, CNRS, Université de Provence, Marseille, France}

\date{\today}

\begin{abstract}
The growth of crystal surfaces, under non-equilibrium conditions,
involves the displacement of mono-atomic steps by atom diffusion and
atom incorporations into steps.  The time-evolution of the growing
crystal surface is thus governed  by a  free boundary value problem
[known as the Burton--Cabrera--Franck model]. In the presence of an
asymmetry of the  kinetic coefficients [Erlich--Schwoebel barriers],
ruling the rates of incorporation of atoms at  each step, it has
been shown that a train of straight steps is unstable to two
dimensional  transverse perturbations. This instability is now known
as the Bales-Zangwill instability (meandering instability). We study
the non-linear evolution of the step meandering instability that
occurs on a crystalline vicinal surface under growth, in the absence
of evaporation, in the limit of  a weak asymmetry of atom
incorporation  at the  steps. We derive a nonlinear amplitude
equation displaying spatiotemporal coarsening. We characterize the
self-similar solutions of this equation.
\end{abstract}


\maketitle

Molecular beam epitaxy (MBE) is often used to grow nano-structures
on vicinal surfaces of semiconductor and metallic crystals
\cite{barabasi95,stangl04,pimpinelli98,saito98,neel03,nita05}. The
ability to grow smooth crystals with a sharp interface  is of
considerable importance when manufacturing electronic and
opto-electronic devices. One possibility for achieving this goal is
to use the step-flow mode, where deposited adatoms diffuse and
attach directly to preexisting steps on a vicinal surface. Ideally
the surface grows without changing its shape by advancement of a
uniform train of step traveling at constant velocity. In
experiments, a wide number of additional effects may modify this
scenario.
Fluctuations in the beam intensity and island growth  may lead to
kinetic roughening of the surface \cite{barabasi95,michely04}. The
presence of impurities can also pin the steps at random positions.
More importantly, intrinsic morphological instabilities such as step
bunching and step meandering can take place on a vicinal surface and
lead to the destruction of a stable train of equidistant straight
steps. Thus, under standard MBE growth conditions (one monolayer per
minute) a rich variety of crystal surface morphologies is
experimentally observed, resulting from the nonlinear evolution of
step bunching and meandering instabilities
\cite{jeong99,politi00,yagi01,neel03}. Moreover, the
self-organization arising from these instabilities has been proposed
as a natural candidate for the development of technological
applications such as quantum dots and quantum wells
\cite{shchukin99,brunner02}.

The step meandering instability was originally predicted
theoretically by Bales and Zangwill \cite{bales90} for a vicinal
surface under growth. Its origin comes from the asymmetry between
the descending and ascending currents of adatoms. As shown by Bales
and Zangwill, a  straight  train of steps during MBE growth may
become morphologically  unstable in the presence of a kinetic
attachment asymmetry at the step: the Erlich-Schwoebel effect (ES).
The physical mechanisms  taking place during this instability are
the destabilizing ES effect combined with the stabilizing effect
provided by the step rigidity. It was shown that the most dangerous
mode is the synchronized mode  for which all the steps have the same
phase \cite{pimpinelli94}. Nonlinear extensions of this work have
shown that the meander evolution can be described by amplitude
equations showing various behaviors. Close to the instability
threshold, starting from the Burton-Cabrera-Frank (BCF) model, it
was shown \cite{bena93} that the step position in the presence of
desorption (evaporation) obeys the Kuramoto-Sivashinsky equation:
\begin{equation}
    \partial_t h= -\partial_{y}^2 h -\partial_{y}^4 h +(\partial_y h)^2
    \, ,
\end{equation}
where $y$ is the coordinate along the step and $x=h(y,t)$ is the
meander amplitude which  describes the step shape in the $(x,y)$
plane. The ultimate stage of this dynamics is thus found to be
spatiotemporal chaos. In  the case of negligible desorption with
strong or moderate ES effect
\cite{pierre-louis98,kallunki00,gillet00}, it was shown that  the
step shape obeys  a highly nonlinear equation
\begin{equation}
    \partial_t h= -\partial_{y}\left[ \frac{1}{1+(\partial_y h)^2}
     \left(\partial_y h +\partial_y (\frac{\partial_{y}^2h}{(1+(\partial_y
     h)^2)^{3/2}})\right)\right] \, .
\end{equation}
This equation  cannot be derived from a weakly nonlinear analysis
but is based on the assumption  that the slope of the steps is order
unity. Instead of  spatiotemporal chaos, a regular pattern is
revealed,  the lateral modulation wavelength is fixed while the
amplitude of the step deformation (transverse meandering amplitude)
increases like $t^{1/2}$. Later, it was shown  that the inclusion of
the elastic step interactions affects the step dynamics in the sense
that they induced a lateral coarsening \cite{paulin01}. Finally, it
was recently shown that interrupted coarsening occurs when
two-dimensional  diffusion anisotropy is included
\cite{danker03,danker04}.

In this paper, we show using standard  weakly nonlinear analysis,
that  under the assumption of negligible desorption and  weak
Erlich-Schwoebel effect (ES),  the time-evolution of the
Bales-Zangwill instability is governed  by the following equation:
\begin{equation}
\partial_t h= -\partial_{y}^2 h -\partial_{y}^4 h +\partial_{y}^2(\partial_y
h)^2 \ .
\end{equation}
 This equation
was already mentioned in Ref. \cite{gillet00} on the basis of
symmetry arguments  as a possible candidates for the time evolution
of  the meandering amplitude but  was not explicitly obtained due to
a different choice of the physical parameters (large and order one
ES effects) \cite{gillet00}. Our results are illustrated by
numerical simulations. We show using a simple similarity argument,
that  the characteristic transverse meandering amplitude (step
width) grows like $t$ and that  the characteristic lateral
coarsening exponent grows like $t^{1/2}$. Our result  are in
agreement with the numerical simulations performed  in Ref.
\cite{kallunki02}.

Let us denote by $x_n(y,t)$  the positions at time $t$ of  the $n$
step (see Fig. 1). For simplicity we neglect elastic interactions
between steps and assume that the desorption of adatoms and
transparency of the steps are negligible. During growth, the adatom
surface concentration on each terrace $C_n(x,y,t)$ , obeys the
following diffusion equation \cite{burton51,schwoebel66,saito98}:
\begin{equation}
\label{diffu1}
 D \nabla^2 C_n(x,y)+F=0 \, ,
\end{equation}
where $D$  is  the adatom diffusion coefficient, and $F$ the
deposition flux. This equation  for   $C_n$ is supplemented by the
following boundary conditions:
\begin{eqnarray}
\label{bcf1}
D \hat{\bm{n}}\cdot\nabla C_n &=&\nu_{+} (C_n-C_{eq,n-1})  \, , \, x=x_{n-1}(y,t) \, ,\\
D \hat{\bm{n}}\cdot\nabla C_n&=&-\nu_{-} (C_n-C_{eq,n}) \,  , \,
x=x_{n}(y,t) \, .
\end{eqnarray}
Here $\nu_{+}$ and  $\nu_{-}$ are the ES coefficients which are
proportional to the rate of attachment of adatoms on the steps from
the terrace; $\hat{\bm{n}}$ is the external normal to the step :
\begin{equation}
\hat{\bm{n}}=(1,-\partial_y x_n)/\sqrt{1+(\partial_y x_n)^2} \, ,
\end{equation}
 and $\nabla$
is the two dimensional gradient operator $(\partial_x,
\partial_y)$. The adatom equilibrium concentrations $C_{eq,n}$
 depend on the step curvatures $\kappa_n$
  \cite{bena93}:
\begin{equation}
\label{courb1}
    C_{eq,n} = C_{(0)}(1+\Gamma \kappa_n)  \, ,
\end{equation}
where $ \Gamma= \Omega \tilde\gamma/k_BT$, with  $\Omega$ the unit
atomic surface, $T$ the temperature,  $k_B$ the Boltzmann constant,
$\tilde \gamma$ the step rigidity and $C_{(0)}$  the reference
adatom concentration. The step curvature is given by
\begin{equation}
\kappa_n=-\frac{\partial_{y}^2 x_n}{(1+(\partial_{y} x_n)^2)^{3/2}}
\, .
\end{equation}
 The normal velocity of  the $n$  step  is given by
\begin{equation}
\label{dyna1} v_n= D \Omega( \hat{\bm{n}}\cdot\nabla
C_{n+1}-\hat{\bm{n}}\cdot\nabla C_{n}) \quad , \quad x=x_{n}(y,t) \,
.\end{equation} The $x$ component of the normal velocities reads:
\begin{equation}
\dot{x}_n=v_n (1+(\partial_y x_n)^2)^{1/2} \, .
\end{equation}
where the dot represents the time derivative.

\begin{figure}
\centering
\includegraphics[width=0.5\textwidth]{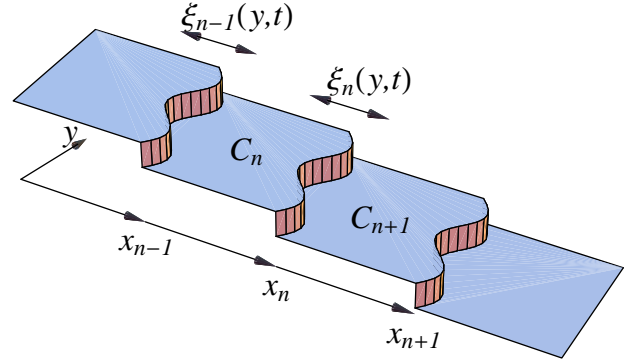}
\caption{Sketch of the  vicinal surface showing a succession of
steps and terraces and the development of the meandering
instability. $x_n$ and $\xi_n(y,t)$
 are step positions and the corresponding
in-phase perturbations. $C_n$ denotes the adatom concentration on
the terrace.} \label{fig1}
\end{figure}

 In order to get a non-dimensional version of equations
(\ref{diffu1},\ref{bcf1},\ref{courb1},\ref{dyna1}), we set the unit
of length to be the initial size of the terrace $l_0$ (initial
 distance between steps) and the unit of time  to be $l_0^3/(C_{(0)} \Gamma
\Omega D)$. Specifically, one sets
\begin{equation}
 \tilde{x}=\frac{x}{l_0} \, , \, \tilde{y}=\frac{y}{l_0}  \, ,  \,
\tilde{C}_n =\frac{l_0[C_n-C_{(0)}]}{C_{(0)}\Gamma}  \, , \,
\tilde{t}=\frac{t\Omega C_{(0)} D\Gamma}{l_0^3}
\end{equation}
in  equations
(\ref{diffu1})-(\ref{bcf1})-(\ref{courb1})-(\ref{dyna1}). We obtain
after omitting the tildes on the variables the following equations
for the dimensionless variables
\begin{eqnarray}
\label{bcf2a}
\nabla^2 C_n&=&-f \\
\label{bcf2b}
\hat{\bm{n}}\cdot\nabla C_n &=&\alpha_+ (C_n-\kappa_{n-1}) \quad , \, x=x_{n-1}(y,t) \\
\label{bcf2c}
\hat{\bm{n}}\cdot\nabla C_n&=&-\alpha_- (C_n-\kappa_{n})  \quad  , \, x=x_{n}(y,t) \\
\label{bcf2d} v_n&=&  \hat{\bm{n}}\cdot\nabla
C_{n+1}-\hat{\bm{n}}\cdot\nabla C_{n}  \, , \, x=x_{n}(y,t)
\end{eqnarray}
The system is thus  controlled by three independent positive
nondimensional parameters:
\begin{equation}
\label{param}
    f=\frac{Fl_0^3}{C_{(0)}\Gamma D}\,,
        \quad
    \alpha_+=\frac{\nu_+ l_0}{D}\,,
        \quad
    \alpha_-=\frac{\nu_- l_0}{D}\,,
\end{equation}
respectively related to the flux and attachment lengths. Let us
investigate the linear stability of a train of equidistant steps
traveling at a constant velocity $f$ when perturbed transversally.
The shape of the steps can be decomposed in Fourier modes of the
form
\begin{equation}
x_n(y,t)=n + f t+\xi_n (y,t)\end{equation}
 where
\begin{equation}
\xi_n(t,y)=e^{\sigma(q,\phi) t+i q y  + i n \phi} \, ,
\end{equation}
 where $q$ and $\phi$  are respectively  the wavenumber and the phase  of the perturbation (see Fig.~1).
Inserting these expressions into equations
(\ref{bcf2a}-\ref{bcf2b}-\ref{bcf2c}-\ref{bcf2d}), we obtain the
general dispersion relation $\sigma=\sigma(q,\phi)$ which possesses,
for each $\phi$, one  maximum \cite{bales90,pimpinelli94}. The
maximum growth rate is reached for the in-phase perturbation
$\phi=0$. Although the full expression of the dispersion relation is
cumbersome, near the instability threshold ( $f=0$), we  introduce a
small parameter $\epsilon$ measuring the distance to the threshold
and we also  assume that  the E-S effect is  small. This latter
assumptions was not used   in   previous works
\cite{pierre-louis98,gillet00}.
 The small  parameter $\epsilon$ arises
naturally when considering the long wavelength limit, in which
$q\rightarrow\epsilon q$. We therefore set
 \begin{equation}
 \label{scale}
    \tilde{f}=\epsilon f \, , \quad
    \tilde{C}=\epsilon^2 C \, , \quad
    \alpha_+= \alpha_- +\epsilon^3 \delta \, , \quad
    \tilde{t}=\frac{t}{\epsilon^2}
\end{equation}
in equations (\ref{bcf2a}-\ref{bcf2b}-\ref{bcf2b}-\ref{bcf2d}).
 This scaling will lead to the following equations after omitting the tilde
 \begin{eqnarray}
\label{bcf3a}
\nabla^2 C_n&=&-f \epsilon \\
\label{bcf3b}
\hat{\bm{n}}\cdot\nabla C_n &=&\alpha_+ (C_n- \epsilon^2\kappa_{n-1}) \quad , \, x=x_{n-1}(y,t) \\
\label{bcf3c}
\hat{\bm{n}}\cdot\nabla C_n&=&-\alpha_- (C_n-\epsilon^2\kappa_{n})  \quad  , \, x=x_{n}(y,t) \\
\label{bcf3d} v_n&=&  \hat{\bm{n}}\cdot\nabla
C_{n+1}-\hat{\bm{n}}\cdot\nabla C_{n}  \, , \, x=x_{n}(y,t) \, .
\end{eqnarray}
 To lowest order in $\epsilon$ we find that the  growth rate
$\sigma \equiv \sigma(q,0)$ is,
\begin{equation}
\label{ss}
    \sigma = \epsilon^6 \left(\delta f \alpha_-  (\alpha_- +2)  \frac{q^2}{2}-  q^4\right) \,, \\
\end{equation}
The growth rate $\sigma$ is maximum for the wavenumber
$q_{{\rm{max}}}=\sqrt{f\delta \alpha_- (2+\alpha_-)/2}$, so that the
most unstable wavelength $\lambda$ should scales $\sim 1/\sqrt{f}$.
The meandering instability originates  from the asymmetry between
the descending and ascending currents of adatoms; the instability
occurs when the rate of attachment of adatoms from the terrace lower
$\alpha_+$ is greater than  the rate of attachment of adatoms from
the upper terrace. In the opposite case, the meandering instability
is not present but a step bunching instability develops. However, at
this moment due to the technical difficulties of the experiments
there has not yet been a clear demonstration of the Bales-Zangwill
instability.

\begin{figure}
\centering
\includegraphics[width=0.4\textwidth]{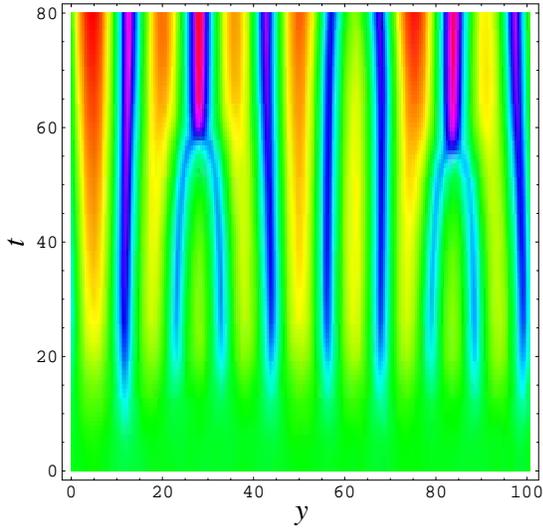}
\caption{Spacetime plot of $u(y,t)$ with nondimensional $y$ and $t$
axes, given by the numerical solution of  equation (\ref{CKS}) with
$\alpha=1$, $\delta=1$ and $f=1$. The coarse-graining of structures
leads to a superposition of parabolas \cite{frisch06}, with a size
$\langle u^2\rangle^{1/2}\sim t$. In the long-time state all the
parabolas tend to have unity curvature at their maximum, and width
increasing as $\sqrt{t}$.} \label{fig3}
\end{figure}

We study  now the nonlinear evolution of the meandering instability,
in the limit of weak amplitudes and long wavelengths. In order to
obtain the relevant nonlinear dynamics we use a standard multi-scale
method. Adatoms concentrations and step shapes are expanded in
powers of $\epsilon$. The adatoms concentrations  depends on the
slowly varying space and time   variables: $Y=\epsilon y$ and
$T=\epsilon^4 t$.  The details of the calculations are deferred to
Appendix A. We thus write
\begin{equation}
\label{delc} C_n(x,y) =\sum_{i=1}^{7} \epsilon^i C_{n}^{(i)}(x,Y)
\quad , \, \xi_n (y,t)=\epsilon u(Y,T) \, .
\end{equation}
Solving diffusion equation (\ref{bcf3a}) and boundary conditions
(\ref{bcf3b}-\ref{bcf3c}) up to order $\epsilon^7$, and inserting
the results into the step velocity equations (\ref{bcf3d}), we find
the equation for the slowly varying amplitude at order $\epsilon^7$.
It may be written as :
\begin{widetext}
\begin{equation}\label{CKS}
    \partial_t u=-\partial_{y}^2\left[
        \frac{\delta f}{2} \alpha  (\alpha +2) u+
        \partial_{y}^2 u+
       \frac{ \mathit{f}  [\alpha  (\alpha +6)+6]}{6\alpha  (\alpha +2) } (\partial_y u)^2
    \right]\, ,
\end{equation}
\end{widetext}
where $\alpha=\alpha_-$.  We have renamed the capital letters $T$
and $Y$ as $t$ and $y$ for simplicity. It is necessary to go to the
seventh order  to balance the nonlinear term with the linear ones.
This equation was recently  presented in the context of surface
growth on a vicinal surface of silicon characterized by different
types of steps and terraces \cite{frisch06}. In this previous work,
it was shown that the anisotropy of diffusion induces a meandering
instability and simple similarity and matching arguments  lead to a
complete picture for the long time behavior of the solution of
equation
 (\ref{CKS}). It is worth noting that Eq. (\ref{CKS}) admits an
exact particular solution in the form of a stationary parabola
\begin{equation}
u(y,t)=-\frac{3 \alpha^2 (2+\alpha)^2 \delta}{4
(6+6\alpha+\alpha^2)}y^2 \, .
\end{equation}
 Trying a similarity solution
$u(y,t)=t^a\varphi(y/t^b)\,$ of equation (\ref{CKS}), we obtained
the exponents $b=1/2$ and $a=1$, which agree with our numerical
results \cite{frisch06}. The general, asymptotic solution of
equation (\ref{CKS}) can be thought of a superposition of parabolas
as shown on the Figure~3 of Ref. \cite{frisch06}. The joining
regions between parabolas can be  matched using  a solution of the
Burgers equation obtained by a Hopf-Cole transformation
\cite{frisch06}. A typical spatiotemporal evolution from a random
initial condition is shown in Fig.~2. The step width (transverse
meandering amplitude) of the steps increase linearly in time
\cite{frisch06} and this results is in agreement with the results
obtained by Kinetic Monte-Carlo methods in the Fig. 7 of Ref.
\cite{kallunki02}.

In this article, we have shown  that the amplitude of the meanders
is governed by Eq. (\ref{CKS}) which displays non-interrupted
coarsening. It would be interesting in a further study to
investigate the effect of phase freedom  of the steps and to derive
a continuous coarse-grained model for the space-time evolution of a
vicinal surface. Such coarse-grained models could also be applied to
step bunching \cite{pierre-louis03,frisch05} and to electromigration
induced instabilities \cite{liu98bis,pierre-louis04}. We hope that
the present work will motivate more experimental research in the
subject and we plan to extent this work to the study of the coupling
between step bunching instabilities and step meandering
instabilities. It would also be interesting to study, using  a
similar line of thought, the effect of an elastic stress commonly
encountered in heteroepitaxy experiments \cite{berbezier02,stangl04}
under step flow conditions \cite{xiang04}. It would also be
interesting to find out if there is a cross-over between the weak
nonlinear regime  we have described in this paper and the strongly
nonlinear regime described in
Refs.~\cite{pierre-louis98,kallunki00,gillet00}.

\acknowledgments{We  thank Olivier Pierre-Louis, Isabelle
Berb\'ezier, Chaouqi Misbah and Mark Mineev for stimulating
discussions.}

\appendix
\section{Details of the expansion}
From equations (\ref{bcf3a}-\ref{bcf3b}-\ref{bcf3c}-\ref{delc}), we
obtain the following relations:
\begin{equation}
\partial_{x}^2 C_n^{(1)}=-f  \, ,
\end{equation}
\begin{equation*}
C_n^{(1)}(x,Y)=\frac{f-f x (x+1) \alpha }{2 \alpha } \, ,
\end{equation*}

\begin{equation}
\partial_{x}^2 C_n^{(2)}=0 \, ,
\end{equation}
\begin{equation*}
C_n^2(x,Y)= \frac{1}{2} f u(Y) (2 x+1) \, ,
\end{equation*}
\begin{equation}
\partial_{x}^2 C_n^{(3)}+\partial_{Y}^2 C_n^{(1)}=0 \, ,
\end{equation}
\begin{equation*}
C_n^{(3)}(x,Y)=-\frac{f u^2(Y)}{2} \, ,
\end{equation*}
\begin{equation}
\partial_{x}^2 C_n^{(4)}+\partial_{Y}^2 C_n^{(2)}=0 \, ,
\end{equation}
\begin{widetext}
\begin{equation*}
C_n^{(4)}(x,Y)=\frac{6 f (x \alpha -1) \delta -f (2 x+1) \alpha ^2
\left[x (x+1) (\alpha +2)-1\right] u''(Y)}{12 \alpha ^2 (\alpha +2)}
\, ,
\end{equation*}
\end{widetext}
\begin{equation}
\partial_{x}^2 C_n^{(5)}+\partial_{Y}^2 C_n^{(3)}=0 \, ,
\end{equation}
\begin{widetext}
\begin{equation*}
C_n^{(5)}(x,Y)= \frac{3 (\alpha +2) \left[f (2 x
   (x+1) \alpha -1) u'(Y)^2-4
   \alpha  u''(Y)\right]+f u
   \left[\alpha  [\alpha +6 x
   (x+1) (\alpha +2)] u''(Y)-6
   \delta \right]}{12 \alpha
   (\alpha +2)} \, ,
\end{equation*}
\end{widetext}

\begin{equation}
\partial_{x}^2 C_n^{(6)}+\partial_{Y}^2 C_n^{(4)}=0 \, ,
\end{equation}

\begin{widetext}
\begin{eqnarray*}
C_n^{(6)}(x,Y)&=& -\frac{f (2 x+1) \left(360 u(Y)
   u'^2(Y) (\alpha +2)^2+180 u^2(Y)
   u''(Y) (\alpha +2)^2\right)}{720 (\alpha
   +2)^2} \\
   & &-\frac{f (2 x+1) \left[
   -\left(\alpha +x
   (x+1) (\alpha +2) (-\alpha +3 x
   (x+1) (\alpha +2)-12)+12\right)
   u''''(Y)\right])}{720 (\alpha
   +2)^2} \, ,
\end{eqnarray*}
\end{widetext}

\begin{equation}
\partial_{x}^2 C_n^{(7)}+\partial_{Y}^2 C_n^{(5)}=0 \, ,
\end{equation}

\begin{equation}
C_m^{(7)}(x,Y)=\frac{1}{(720 \alpha ^3 (\alpha +2)^2)} \gamma \, .
\end{equation}

\begin{widetext}
\begin{eqnarray*}
\gamma &= & 360 f (\alpha +2)^2 u^2(Y) u'^2(Y) \alpha ^3+120
   f (\alpha +2)^2 u^3(Y) u''(Y) \alpha ^3 \\
   && -f
   \left(30 (\alpha +2)^2 x^4+60 (\alpha +2)^2
   x^3+30 \alpha  (\alpha +2) x^2-60 (\alpha +2)
   x-\alpha  (\alpha +12)\right) u(Y) u''''(Y)
   \alpha ^3 \\
   &&-30 \left(f \alpha  (\alpha +2)
   (\alpha  (2 (\alpha +6)+x (x+1) (\alpha  (3 x
   (x+1) (\alpha +2)-2 (\alpha +6))-12))+12)
   u''^2(Y)\right) \\
   && -30 \left(-2 f \alpha  \left(\alpha  \left(3
   (\alpha +2) x^2+2 (\alpha +3)
   x-2\right)-6\right) \delta  u''(Y)\right) \\
   && -30 \left(2 \left(6
   f (\alpha +1) (x \alpha -1) \delta ^2\right)\right)\\
   && -30\left(+\alpha
   (\alpha +2) \left(f (\alpha  (\alpha +x (x+1)
   (\alpha  (-\alpha +2 x (x+1) (\alpha
   +2)-7)-6)+6)+6) u'(Y) u'''(Y)\right)\right) \\
   &&-30\left(-6 \alpha
   (\alpha +2) (x (x+1) \alpha -1)
   u''''(Y)\right) \, ,
\end{eqnarray*}
\end{widetext}

Here the symbol  $'$ denotes the derivative with respect to  the
variable  $Y$.

\bibliography{mineev1}

\end{document}